\documentclass[a4paper,prl,showpacs,superscriptaddress,twocolumn]{revtex4}

\usepackage{amsfonts}
\usepackage{amssymb}
\usepackage{amsmath}
\usepackage{calc}
\usepackage{graphicx}
\usepackage{array}
\usepackage{bm}
\usepackage{delarray}
\usepackage{ifthen}

\begin{document}
\title{Engineering mixed states in a degenerate four-state system}

\author{A.  Karpati} 
\affiliation{H.A.S.  Research Institute for Solid
  State  Physics and  Optics,  H-1525 Budapest,  P.O.Box 49,  Hungary}
\author{Z.  Kis}  
\affiliation{H.A.S.   Research Institute  for  Solid
  State  Physics and  Optics,  H-1525 Budapest,  P.O.Box 49,  Hungary}
\author{P.   Adam} 
\affiliation{H.A.S.  Research  Institute for  Solid
  State  Physics and  Optics,  H-1525 Budapest,  P.O.Box 49,  Hungary}
\affiliation{H.A.S.  Research  Group   for  Nonlinear  and  Quantum
  Optics, and Institute of  Physics, University of P\'ecs, Ifj\'us\'ag
  \'ut 6., H-7624 P\'ecs, Hungary}
 
\begin{abstract}
  A method  is proposed  for preparing  any pure and  a wide  class of
  mixed quantum  states in the  decoherence-free ground-state subspace
  of  a  degenerate  multilevel   lambda  system.   The  scheme  is  a
  combination of  optical pumping and a series  of coherent excitation
  processes, and  for a given pulse  sequence the same  final state is
  obtained regardless of the initial  state of the system.  The method
  is robust with  respect to the fluctuation of  the pulse areas, like
  in adiabatic  methods, however, the field amplitude  can be adjusted
  in a larger range.
\end{abstract}

\date{\today }


\maketitle

Controlling the quantum state of degenerate quantum systems have drown
much attention  recently.  This  field has developed  independently in
several  different series  of  studies: Among  the numerous  adiabatic
passage techniques \cite{Vitanov01} one  of the most well-known method
is the stimulated Raman  adiabatic passage (STIRAP) \cite{AAMOP}.  The
STIRAP can  be used not  only for transferring population  between two
quantum states using crafted laser pulses, but it has been utilized to
create  coherent  superpositions  in  three-  and  four-level  systems
\cite{Marte91,  Unanyan98, Theuer99},  to  prepare maximally  coherent
superposition   states   \cite{Unanyan01}   and   arbitrary   coherent
superpositions  \cite{Kis01,   Kis02,Kis03}  in  $N$-state  degenerate
systems.  The  applicability  of  the  STIRAP  method  is  limited  by
constraints on the field amplitudes \cite{AAMOP,Vitanov97}.

The  other  field  that   developed  toward  the  quantum  control  of
degenerate systems  is termed  ``coherent control'' that  uses several
interfering  pathways in  the quantum  system to  transfer selectively
population from  an initial  state to a  target one  \cite{Shapiro03}. 
Merging this  technique with the STIRAP  method led to  the mapping of
wave-packets between  vibrational potential surfaces  in molecules for
the      non-degenerate       \cite{Kraal02a}      and      degenerate
\cite{Thanopulos04,Gong04} cases.

The above  mentioned control processes  have great importance  in many
areas  of  quantum-information  processing  (QIP),  involving  quantum
computing,   cryptography  and  teleportation   \cite{Nielsen00}.   In
general,  mixed   states  cannot   be  created  with   coherent  state
preparation methods. However, for several QIP problems it is essential
to develop  quantum state preparation techniques which  are capable to
prepare  not  only  pure, but  mixed  states  of  the system  as  well
\cite{Bacon01,Somma02,Tarasov02}.

\begin{figure}
   \includegraphics[width=4cm]{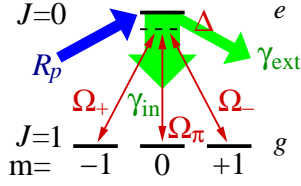}
   \caption{(Color Online) The coupling scheme for our state-engineering 
     procedure: the lower states are the magnetic sublevels of a $J=1$
     angular momentum  state which  are coupled by  $\sigma_{\pm}$ and
     $\pi$ polarized  pulses to a  single excited state.   The excited
     state decays with a rate $\gamma_{\rm in}$ into the lower states,
     and  it may  decay out  of the  system with  a  rate $\gamma_{\rm
       ext}$. For non-zero $\gamma_{\rm  ext}$ a repumping is switched
     on with a rate $R_p$. }
\label{fig:scheme}
\end{figure}

In  optical pumping  processes  \cite{pump}, the  final  state of  the
system  is largely  independent  of its  initial  state, however,  the
efficiency  is  small  \cite{Shore90}.   On  the other  hand,  in  the
coherent  state-preparation methods  the  final state  depends on  the
initial state  of the system, but  the efficiency can  be nearly unity
\cite{Vitanov01,AAMOP}.  In  this letter  we consider a  novel concept
for  quantum-state  preparation, which  is  a  combination of  optical
pumping  and  coherent   excitation  processes,  exhibiting  only  the
advantageous properties of the two schemes, and capable to prepare not
only pure but  prescribed mixed states of the  system too.  The unique
features of our method compared to other state-preparation methods are
the following: it  is simultaneously (i) robust, (ii)  the final state
is independent of the initial state  of the system, (iii) the state is
prepared  in  a decoherence-free  subspace,  (iv)  the  choice of  the
excitation  field amplitude  is quite  arbitrary.  The  method  can be
implemented in  multilevel lambda  systems.  For concreteness,  let us
consider the  four-state system shown  in Fig.~\ref{fig:scheme}: there
are three degenerate ground states  and a single excited state coupled
by an elliptically polarized  coherent laser pulse.  The ground states
$|g_q\rangle$ $(q=-, \pi, +)$ are assumed to be the magnetic sublevels
of  a  $J_g=1$  angular  momentum  state, whereas  the  excited  state
$|e\rangle$  has $J_e=0$.   The three  polarization components  of the
coupling field, denoted by ${\cal E}_q$ with $(q=-, \pi, +)$, 
share the same time-dependence, but they can have different
peak amplitudes and phases
\begin{subequations}  \label{Rabis}
\begin{eqnarray}
  {\cal E}_-(t) &=& {\cal E}(t)\,e^{i\xi} e^{i\mu_-}\sin\theta\sin\varphi\,,\\
  {\cal E}_\pi(t) &=&  {\cal E}(t) e^{i\xi} \cos\theta\,,\\
  {\cal E}_+(t) &=& {\cal E}(t)\,e^{i\xi} e^{i\mu_+}\sin\theta\cos\varphi\,,
\end{eqnarray}
\end{subequations}
where the parameters $\theta$,  $\varphi$ describe the polarization of
the pulses, $\xi$  is the absolute phase of the  pulse, and the phases
of  the $\sigma_+$  and $\sigma_-$  components relative  to  the $\pi$
component are  $\mu_+$ and  $\mu_-$, respectively.  The  excited state
$|e\rangle$ decays with spontaneous emission to the ground states with
a rate of $\gamma_\textrm{in}$, and  it may decay to states other than
the three ground states with  a rate of $\gamma_\textrm{ext}$ as well. 
When decay out  of the ground-state space occurs,  a repumping process
with a  rate of $R_p$  is switched on  in order to compensate  for the
population loss.   We show that  by using a predetermined  sequence of
pulses we  can create  any prescribed  pure or a  wide class  of mixed
final  states  in the  ground-state  space,  starting  from {\em  any}
initial state.

The  Master equation describing  the time evolution  of the  system is
given by
\begin{eqnarray}\label{eq:master}
  \frac{\mathrm d}{\mathrm d t}\widehat\varrho&=&\frac1{i\hbar}
  [\widehat H,\widehat\varrho]\nonumber\\
  &+&\frac{\gamma_\mathrm{in}}2\sum_{l=-, \pi, +}
  2\widehat L_l^{\phantom{\dagger}}\widehat \varrho\widehat L_l^\dagger - 
  \widehat L_l^\dagger \widehat L_l^{\phantom{\dagger}}\widehat \varrho
  - \widehat\varrho \widehat L_l^\dagger \widehat L_l^{\phantom{\dagger}} \\
  &-&\frac{\gamma_\textrm{ext}}{2} \widehat L_e \widehat\varrho - 
  \frac{\gamma_\textrm{ext}}{2}  \widehat\varrho \widehat L_e+
  R_p (1-{\mathrm Tr} \left\{\widehat\varrho\right\}) \widehat L_e,\nonumber
\end{eqnarray}
where the Hamiltonian $\widehat H$ reads
\begin{eqnarray}
  \widehat H &=& \frac{\hbar}{2}( \Omega_- |g_-\rangle\langle e|\, +\,
  \Omega_\pi |g_\pi\rangle\langle e|\, +\, \Omega_+ |g_+\rangle\langle e|+
  \textrm{h.c.})
  \nonumber\\
  & +& \hbar\Delta|e\rangle\langle e| \,,
  \label{ham}
\end{eqnarray}
where the  Rabi frequencies are  $\Omega_- = \frac{1}{3} \Omega
e^{i\xi}    e^{i\mu_-}   \sin\theta   \sin\varphi$,    $\Omega_\pi   =
-\frac{1} {3}   \Omega   e^{i\xi}  \cos\theta$,   $\Omega_+   =
\frac{1}{3} \Omega  e^{i\xi} e^{i\mu_+} \sin\theta \cos\varphi$,
with $\hbar\Omega=d_{ge}{\cal E}$.  The  step operators are defined as
$  \widehat   L_q  =  \frac{1}{\sqrt 3}   |g_q\rangle\langle  e|$  and
$\widehat L_e  = |e\rangle\langle e|$.  The other  symbols are defined
above.

The  Hamiltonian  of  Eq.~(\ref{ham})  has two  uncoupled  eigenstates
$|\Phi_D^{(l)}\rangle$ \cite{Morris83},  i.e. they are  decoupled from
the  external driving  field,  $\widehat H|\Phi_D^{(l)}\rangle=0$  for
$l=1, 2$.  They read
\begin{equation}\label{dstate}
  |\Phi_D^{(l)}\rangle =\sum_{q=-, \pi, +} n_q^{(l)} |g_q\rangle\,, 
  \quad l=1, 2\,,
\end{equation}
where   the  unit  vectors   ${\bm  n}^{(l)}$   are  given   by  ${\bm
  n}^{(1)}=[e^{ i\mu_-}  \cos\theta\sin\varphi, \sin\theta, e^{i\mu_+}
\cos\theta       \cos\varphi]^T$       and      ${\bm       n}^{(2)}=[
-e^{-i\mu_+}\cos\varphi, 0,  e^{-i\mu_-}\sin\varphi]^T$, and the field
parameters   $\mu_\pm$,    $\varphi$,   $\theta$   are    defined   in
Eq.~(\ref{Rabis}).  These states are  dark states, because they do not
have  a  component  in   the  excited  state  \cite{Arimondo96}.   The
Hamiltonian has two other  eigenstates with non-zero eigenvalues, they
are  called bright states,  because they  do have  a component  in the
excited state \cite{Arimondo96}.

Let    us    assume    that     we    have    some    initial    state
$\widehat\varrho_\textrm{in}$   defined   in   the   decoherence-free,
ground-state  space.   The Hamiltonian  part  of  the Master  equation
(\ref{eq:master}) drives  the bright components  of this state  to the
excited state  back and forth  via Rabi oscillations.  As  the excited
state  gets  populated the  spontaneous  emission  will interrupt  the
Hamiltonian dynamics  and the system falls back  into the ground-state
space or some other external states become populated. As a result, the
two {\em dark states} become more and more populated, even though they
are decoupled from the external driving field.  When there is no decay
into  external states or  the decay  out of  the four-state  system is
compensated by an incoherent  repumping process, the system will relax
into  the dark subspace  of the  Hamiltonian. Then,  the state  of the
system is given by
\begin{equation}
  \widehat\varrho_\textrm{out}=p^{(1)}|\Phi_D^{(1)}\rangle\langle\Phi_D^{(1)}|
  +p^{(2)}|\Phi_D^{(2)}\rangle\langle\Phi_D^{(2)}|\,,
\end{equation}
where the coefficients  $p^{(l)}$ depend on the applied  pulse and the
initial state as well. It is  important to note that this output state
is independent of the pulse amplitude $\cal E$, it depends only on the
polarization and relative phases of  the three components of the field
of Eq.~(\ref{Rabis}).

In   the  ground-state   space   the  state   $|\Phi^{(\perp)}\rangle$
orthogonal   to    the   dark   states    of   Eq.~(\ref{dstate})   is
$|\Phi^{(\perp)}\rangle   =    e^{   i\mu_-}   \sin\theta\sin\varphi\,
|g_-\rangle   -\cos\theta\,  |g_\pi\rangle  +   e^{i\mu_+}  \sin\theta
\cos\varphi \,  |g_+\rangle$\,.  This vector can point  to anywhere in
the three-dimensional ground-state space, depending on the laser-field
parameters.  Consequently,  it is  possible to choose  the laser-field
parameters,  so  that  any  two  linearly  independent  state  vectors
$|\psi_1\rangle$   and  $|\psi_2\rangle$   of   the  three-dimensional
ground-state space lay in  the dark subspace.  Therefore, in principle
there exists a pulse-sequence, such that any prescribed final state of
the form
\begin{equation}\label{rhof}
  \widehat\varrho_\textrm{f}=p_\textrm{f}^{(1)}|\psi_1\rangle\langle\psi_1|
  +p_\textrm{f}^{(2)}|\psi_2\rangle\langle\psi_2|\,,
\end{equation}
can be obtained.  The state $\widehat\varrho_\textrm{f}$ can be either
a pure state if one of the coefficients $p_\textrm{f}^{(l)}$ vanishes,
or a mixed state if both of them are non-zero.

We  have   two  tasks   now:  (i)  to   find  how  an   initial  state
$\widehat\varrho_\textrm{in}$     transforms    when     the    pulses
Eq.~(\ref{Rabis}) are  adjusted to a  certain value; (ii) to  find the
pulse-sequence  that steer  the state  of the  system to  a prescribed
final state defined by Eq.~(\ref{rhof}).

For   convenience,  the   linear  space   of  the   density  operators
$\{\widehat\varrho\}$  is represented  by vectors  $\{{\bm  r}\}$ with
components $  r_{4(i\smash{-}1) \smash{+}j}= (\widehat\varrho)_{i,j}$,
where $(\widehat\varrho)_{i,j}$  is the matrix element  of the density
operator  $\widehat\varrho$  in   the  ordered  basis  $\{|g_-\rangle,
|g_\pi\rangle,  |g_+\rangle, |e\rangle\}$\,.   The  scalar product  of
vectors   is    defined   as   $({\bm    r}^{(1)}|{\bm   r}^{(2)})   =
\sum_s{r^{(1)*}_s}r^{(2)}_s  =  {\rm Tr}  \left\{\widehat\varrho^{(1)}
  \widehat\varrho^{(2)}\right\}$.        The      Master      equation
(\ref{eq:master}) in this representation takes the form $\frac{\mathrm
  d}{\mathrm d t}\bm r=\bm M\bm r+  \bm d$, where the matrix ${\bm M}$
describes the  linear part  of the Master  equation (\ref{eq:master}),
and ${\bm  d}$ corresponds to  the constant term $R_p\widehat  L_e$ in
the incoherent repumping  of the excited state. In  this letter we are
going to consider two cases:

{\em ($\alpha$)\quad  The case $\gamma_\textrm{ext}=R_p=0$  :} In this
case the  Master equation is  homogeneous in $\bm  r$, and $\bm  d$ is
zero.   The relaxation  of  the system  into  its final  state can  be
described by those left- and right-hand eigenvectors (denoted by ${\bm
  r}_{\rm L}^{(k)}$ and ${\bm  r}_{\rm R}^{(k)}$, respectively) of the
matrix $\bm M$, which belong to the eigenvalue zero
\begin{equation}
  \bm M\bm r^{(k)}_{\rm R}=\bm 0\,,\quad
  {\bm r^{(k)\,T}_{\rm L}} \bm M=\bm 0\,,
\end{equation}
and   are   orthonormal   $(\bm   r^{(k)}_{\rm   L}|\bm   r^{(l)}_{\rm
  R})=\delta_{kl}$.  The left- and  right-hand zero subspaces of ${\bm
  M}$  are  four-dimensional  and  they are  different.   The  density
matrices   corresponding   to   the   right-hand   eigenstates   ${\bm
  r}_\textrm{R}^{(k)}$   are  composed   from  the   dark  eigenstates
$|\Phi_D^{(l)}\rangle$ of the Hamiltonian (\ref{ham}), as
\begin{subequations}\label{rhoeig}
\begin{eqnarray}
  \widehat \varrho^{(1)}_{\rm R}&=&\frac 1{\sqrt{2}}\left(
    |\Phi_D^{(1)}\rangle\langle\Phi_D^{(1)}|-|\Phi_D^{(2)}\rangle\langle\Phi_D^{(2)}|
  \right)\,,  \label{rhoe1}\\
  \widehat \varrho^{(2)}_{\rm R}&=&\frac1{\sqrt{2}}\left(|\Phi_D^{(1)}\rangle 
    \langle\Phi_D^{(2)}|+|\Phi_D^{(2)}\rangle\langle\Phi_D^{(1)}|\right)\,,
  \label{rhoe2}\\
  \widehat \varrho^{(3)}_{\rm R}&=&\frac{i}{\sqrt{2}}\left(|\Phi_D^{(2)}\rangle
    \langle\Phi_D^{(1)}|-|\Phi_D^{(1)}\rangle\langle\Phi_D^{(2)}|\right)  \,,
  \label{rhoe3}\\
  \widehat \varrho^{(4)}_{\rm R}&=&\frac 1{\sqrt{2}}\left(
  |\Phi_D^{(1)}\rangle\langle\Phi_D^{(1)}|+|\Phi_D^{(2)}\rangle\langle\Phi_D^{(2)}|
  \right)\,.
\end{eqnarray}
\end{subequations}
As for the density  matrix representation of the left-hand eigenstates
${\bm     r}_\textrm{L}^{(k)}$,    the    first     three    $\widehat
\varrho^{(k)}_{\rm   L}$   are    given   by   Eqs.~(\ref{rhoe1})   --
(\ref{rhoe3}),  and  tho fourth  one  is $\widehat  \varrho^{(4)}_{\rm
  L}=\frac1 {\sqrt 2}  \hat I$.  The final state  of the system, after
the  relaxation  has finished,  is  given by  $  {\bm  r}_{\rm out}  =
\sum_{k=1}^4   (\bm  r^{(k)}_{\rm  L}   |  \bm   r_{\rm  in}   )  {\bm
  r}^{(k)}_{\rm R}  $ since in this state  $\frac{\mathrm d}{\mathrm d
  t}{\bm  r}_\textrm{out}=\bm  0$.   By  using  the  density  operator
representation   Eq.~(\ref{rhoeig})    of   the   eigenvectors   ${\bm
  r}^{(k)}_{\rm L/R}$ of $\bm  M$, the input-output transformation can
be written in a simple  form as $\widehat\varrho_\mathrm{ out} = {\cal
  T}_a      (\widehat     \varrho_\mathrm{in})$,      where     ${\cal
  T}_a(\widehat\varrho)$ reads
\begin{equation}
  {\cal T}_a(\widehat\varrho) =\widehat\varrho'+
  \frac12(1-{\rm Tr}\left\{\widehat\varrho'\right\})  \widehat P_D\,,
  \quad
  \widehat\varrho'=\widehat P_D  \widehat\varrho \widehat P_D
  \label{rhomap1}
\end{equation}
where  $\widehat P_D$ is  a projector  into the  dark subspace  of the
Hamiltonian      (\ref{ham}),     $      \widehat     P_D=\sum_{k=1}^2
|\Phi_D^{(k)}\rangle \langle\Phi_D^{(k)}| $\,.

{\em $(\beta)$  \quad The case $R_p,  \gamma_\textrm{ext}>0$:} Now the
excited   state  is   repumped  from   all  external   decay  channels
incoherently with  a rate of $R_p$.  The  linear differential equation
that governs the time evolution of the density operator takes the form
$  \frac{\mathrm d}{\mathrm  d  t}({\bm r}  - \widetilde{\bm  r})={\bm
  M}'({\bm  r}  -  \widetilde{\bm  r})$,  where  the  constant  vector
$\widetilde{\bm r}$  satisfies ${\bm M}' \widetilde{\bm r}  = -\bm d$,
and the density matrix corresponding to $\widetilde{\bm r}$ is
\begin{eqnarray}\label{rhotilde}
  \widehat{\widetilde{\varrho}} &=& 
  \sin^2\varphi|g_+\rangle\langle g_+|\, +\, 
  \cos^2\varphi |g_-\rangle\langle g_-|\nonumber\\
  &-&\frac12(e^{i(\mu_+-\mu_-)}
  \sin2\varphi |g_+\rangle \langle g_-|   \, +\,\textrm{ h.c.})\,.
\end{eqnarray}
The  left- and  right-hand  zero-subspaces of  the  matrix ${\bm  M}'$
coincide  and  they  are  three-dimensional.  The  eigenvectors  ${\bm
  r}^{(i)}$, ($i=1,2,3$) satisfy  the equation ${\bm M}'({\bm r}^{(i)}
- \widetilde{\bm  r})=\bm 0$, and  the corresponding  density matrices
are  given  by   Eqs.~(\ref{rhoe1})--(\ref{rhoe3}).   Instead  of  the
mapping  in the  case ({\em  $\alpha$}), the  input-output  states are
connected through the relation $ \bm r_{\rm out} = \widetilde{\bm r} +
\sum_{i=1}^3  ( \bm  r^{(i)}  | \bm  r_{\rm in}-\widetilde{\bm  r})\bm
r^{(i)},$   which   in  the   density   matrix  representation   reads
$\widehat\varrho_\mathrm{     out}    =    {\cal     T}_b    (\widehat
\varrho_\mathrm{in})$, where  ${\cal T}_b(\widehat\varrho)$ is defined
as
\begin{equation}
  {\cal T}_b(\widehat\varrho) =\widehat{\widetilde{\varrho}}
  -\widehat{\widetilde{\varrho}}'+   \widehat{\varrho}'+
  \frac12(1-{\rm Tr}\left\{\widehat\varrho'\right\})  \widehat P_D\,,
  \label{rhomap2}
\end{equation}
where  the prime  denotes  projection  into the  dark  subspace as  in
Eq.~(\ref{rhomap1}).

Now we  turn our attention to  finding a pulse sequence  that yields a
desired  final density  operator  of the  form Eq.~(\ref{rhof}).   The
transformation  of an  initial density  operator is  described  by the
subsequent  applications of  the mappings  of Eqs.  (\ref{rhomap1}) or
(\ref{rhomap2})
\begin{equation}
  \widehat{\overline\varrho}_{\rm f}={\cal T}^{(N)}
  (\,{\cal T}^{(N-1)}(\hdots {\cal T}^{(1)}
  (\widehat\rho_{\rm i})\hdots ))\,,
  \label{rhoopt}
\end{equation}
where    ${\cal    T}(\widehat\varrho)$    is    equal    to    ${\cal
  T}_a(\widehat\varrho)$ or ${\cal T}_b(\widehat\varrho)$.  We have to
choose  the number  of  steps $N$,  then  to find  the relative  pulse
amplitudes  and  phases, defined  in  Eq.~(\ref{Rabis}),  by means  of
minimizing numerically the functional
\begin{eqnarray}
  {\cal J}(\{{\cal{E}}\},\widehat\varrho_{\rm in}, 
  \widehat\varrho_{\rm f}) = \left(1-{\rm Tr}\{
  \widehat{\overline\varrho}_{\rm f}
  \widehat{\varrho}_{\rm f}\}\right)^{1/2}\,,
  \label{optim}
\end{eqnarray}
which      is     the      mismatch      between     the      obtained
$\widehat{\overline\varrho}_{\rm  f}$\,  (Eq.~(\ref{rhoopt})) and  the
required   $\widehat{\varrho}_{\rm   f}$\,  (Eq.~(\ref{rhof}))   final
density  operators.  The  numerical optimization  can be  performed by
means of e.g. the conjugate gradient method \cite{numrec}.  Due to the
special linear property of the mappings ${\cal T} (p_1\widehat\rho_1 +
p_2\widehat\rho_2)=p_1{\cal   T}   (\widehat\rho_1)   +  p_2{\cal   T}
(\widehat\rho_2)$  for  $p_1+p_2=1$, it  is  sufficient  to study  the
convergence for  pure initial states,  which are the  arbitrary linear
superpositions of the ground states.  Our aim is to reach a prescribed
destination density  operator by applying  the {\em same  fixed} laser
pulse sequence for all initial states.

Let us consider a concrete example to demonstrate the efficiency of the
proposed  state engineering  method: We  choose the destination density
operator as
\begin{equation}
  \widehat{\overline\varrho}_{\rm f} = \frac 13 |\psi_{\rm f}^{(1)}\rangle
  \langle\psi_{\rm f}^{(1)}| +
  \frac 23 |\psi_{\rm f}^{(2)}\rangle\langle\psi_{\rm f}^{(2)}|,
  \label{rhod}
\end{equation}
with two pure states
\begin{equation}
|\psi_{\rm f}^{(1)}\rangle=\left[\begin{array}{c}
\frac27 e^{i\pi/3}\\[2pt]
\frac37e^{i\pi/5}\\[2pt]
\frac67\\[2pt]
0
\end{array}\right]\,,\quad
|\psi_{\rm f}^{(2)}\rangle=\left[\begin{array}{c}
\frac35\\[2pt]
\frac45e^{i\pi/7}\\[2pt]
0\\[2pt]
0
\end{array}\right]\,.
\end{equation}

\begin{figure}
   \includegraphics[width=\textwidth/2-0.3cm]{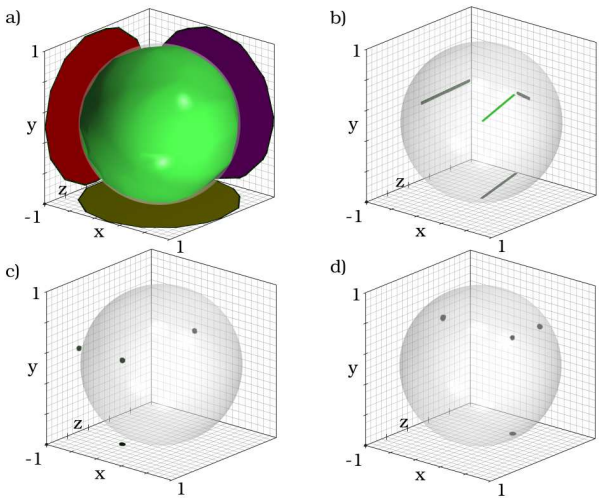}
   \caption{(Color Online) The transformation of the initial-state set 
     after the  first- (a), second-  (b), third- (c), and  fourth- (d)
     excitation   steps,   shown    in   the   Bloch-sphere   of   the
     two-dimensional  dark  subspace.   The  coordinates  are  defined
     through                        the                       relation
     $\widehat\varrho=\frac12(1+x\widehat\sigma_x+y\widehat\sigma_y+
     z\widehat\sigma_z)$,  where  $\widehat\sigma_q$  are the  Pauli's
     spin  operators.  The projections  of  the  distributions to  the
     coordinate planes are also shown. }
\label{fig:sph1}
\end{figure}

First we  discuss the case when $\gamma_{\rm  ext}=R_p=0$: The initial
set   $\overline{\cal   H}$    is   obtained   by   discretizing   the
four-dimensional  parameter  space  --  two relative  phases  and  two
relative amplitudes  -- describing the  possible pure initial  states. 
Then  we have  taken a  four-step excitation  process, i.e.   $N=4$ in
Eq.~(\ref{rhoopt}), and used the conjugate gradient method to minimize
the   functional  ${\cal  J}(\{{\cal   E}\},\widehat\varrho_{\rm  in},
\widehat\varrho_{\rm   f})$  of   Eq.~(\ref{optim})   on  the   subset
$\overline{\cal      H}$,     $\widehat\varrho_{\rm     in}=|\psi_{\rm
  in}\rangle\langle  \psi_{\rm  in}|$  and  $|\psi_{\rm  in}\rangle\in
\overline{\cal H}$.  The outcome of  the optimization is a sequence of
four   polarization  angles   and  relative   phases  ($\varphi^{(l)},
\theta^{(l)},  \mu_-^{(l)}, \mu_+^{(l)}$)  for $l=1,\ldots  ,4$, which
characterize the  pulse sequence $\{{\cal E}\}$.   This pulse sequence
effects  for any  initial  state such  a  final state,  for which  the
mismatch Eq.~(\ref{optim}) is less  than $\approx 10^{-5}$ (limited by
machine-precision).   The subsequent stages  of the  transformation of
the initial set $\overline{\cal  H}$ are shown in Fig.~\ref{fig:sph1}. 
After  the  first  pulse  (Fig.~\ref{fig:sph1}a) the  closure  of  the
transformed  initial set  is the  surface  of the  Bloch sphere.   The
second  step (Fig.~\ref{fig:sph1}b)  yields an  elongated cigar-shape,
while the third step an ellipsoid (Fig.~\ref{fig:sph1}c) distribution,
finally   the  fourth   step   (Fig.~\ref{fig:sph1}d)  contracts   the
distribution to  a point-like region  in the Bloch sphere  with radius
around  $\approx  10^{-5}$.  Then  we  solved  numerically the  Master
equation  (\ref{eq:master})  inserting   the  obtained  optimal  pulse
sequence,  with a  constant $\Omega=1$  and $\gamma_{\rm  in}=1$.  The
pulse duration for  each step should be chosen  so that the relaxation
process into  the actual dark subspace  terminates practically.  These
times can be estimated from the eigenvalues of the matrix $\bm M$: the
one  with the  smallest absolute  real part  limits the  speed  of the
convergence.
We  note that  we  have found
convergence for any other prescribed final state as well, however, the
required number  of steps depends on  the purity of  the target state:
the  purer the state  (i.e.  $p_{\rm  f}^{(1)}\ll p_{\rm  f}^{(2)}$ or
$p_{\rm f}^{(1)}\gg p_{\rm f}^{(2)}$ in Eq.~(\ref{rhof}) ), the larger
the number of steps required.

For  non-zero $\gamma_{\rm  ext}$ and  $R_p$ the  optimization  of the
pulse sequence  can be done  as in the  previous case. In  a four-step
process, the  numeric optimization yielded a pulse  sequence for which
the mismatch  (Eq.~(\ref{optim})) is less than  $\approx 10^{-5}$.  We
have found  that the shape  of the initial distribution  transforms in
the course of  the subsequent stages of the  excitation process in the
same manner as before.  Then we solved numerically the Master equation
using the obtained optimal  pulse sequence, setting the Rabi frequency
$\Omega$,  the  decay  constants  $\gamma_{\rm in}$  and  $\gamma_{\rm
  ext}$, and the repumping rate $R_p$ to unity: we have found that the
process converges similarly to the previous case.
We note  that the ratios  of $\gamma_{\rm in}, \gamma_{\rm  ext}$, and
$R_p$  influence the  rapidity of  the convergence.   For sufficiently
long time steps the process always converges.

In summary we  have worked out a  scheme to create any pure  or a wide
class of  mixed states  in a four-state  degenerate $\Lambda$  system. 
Our method  is based on an excitation-relaxation  process, that drives
the state of the system into the dark subspace of the Hamiltonian that
governs the dynamics without the decay processes.  Although our method
is not adiabatic, it is robust, because the final state is insensitive
to the fluctuations  in the pulse-area of the  applied laser-field.  A
particular advantage of the  method compared to the adiabatic schemes,
that here we have a larger freedom to choose the field amplitude.

The  authors acknowledge  the support  from the  Research Fund  of the
Hungarian  Academy  of Sciences  (OTKA)  under  contracts T043287  and
T034484.  ZK acknowledges the  support from the J\'anos Bolyai program
of the Hungarian Academy of Sciences.

\end{document}